\RequirePackage{fix-cm}
\documentclass[twocolumn,epjc3]{svjour3}  
\smartqed  
\RequirePackage{graphicx}
\journalname{Eur. Phys. J. C}
\begin{document}

\title{Perturbative Resonance during Dark Matter Production and Its Cosmological Implications}

\author{Changhong Li\thanksref{e1, addr1}
}

\thankstext{e1}{e-mail: changhongli@ynu.edu.cn}


\institute{Department of Astronomy, Key Laboratory of Astroparticle Physics of Yunnan Province,  School of Physics and Astronomy,  Yunnan University, No.2 Cuihu North Road, Kunming,  650091 China \label{addr1}
}

\date{Received: date / Accepted: date}

\maketitle

\begin{abstract}
To trace the particle nature of dark matter (DM) from cosmic microwave background observations, we study the co-evolution of DM density perturbation and primordial metric perturbation during DM production.  Solving the perturbative Einstein and Boltzmann equations, we unveil an amplified perturbative resonance between the DM density perturbation and the scalar modes of metric perturbation, driven by DM production. Such perturbative resonance does not affect the tensor modes and suppresses the tensor-to-scalar ratio of metric perturbation $r$.  Using a typical model to specify cosmic reheating, we obtain a direct relation between DM particle mass $m_\chi$ and $r$, which can constrain $m_\chi$ in future primordial gravitational wave searches.
\end{abstract}

\section{Introduction}
\label{intro}
Dark matter (DM), a non-luminous matter component beyond the Standard Model (SM) of Particle Physics which comprises nearly $80\%$ of the matter in the current Universe \cite{WMAP:2010qai}\cite{Planck:2015fie}, plays a crucial role in the formation of structures from cosmic scale to sub-galactic scale at the dawn of our Universe \cite{Frenk:2012ph}.  Unveiling the particle nature of DM, therefore, is one of the leading topics in physics and astronomy.  For existing DM models such as the weakly interacting massive particles (WIMP) paradigm \cite{Lee:1977ua}  and the non-equilibrium thermal DM (NEQDM) model \cite{Li:2014era}, using the DM density fraction, $\Omega_\chi=0.26$ \cite{Bertone:2004pz}, one can obtain the relation between DM particle mass $m_\chi$ and its thermally averaged cross-section $\widetilde{\langle \sigma v\rangle}$.  However, current evidence drawn from galaxies, clusters, cosmic microwave background (CMB) anisotropy, etc., can determine neither $m_\chi$ nor  $\widetilde{\langle \sigma v\rangle}$ separately, as such evidence reflects only the macroscopic gravitational effects of DM rather than its elementary particle nature  \cite{Bertone:2016nfn}.  Lacking conclusive evidence of a single DM particle also leads to a long-standing controversy between cold DM models and other exotic hypotheses \cite{ParticleDataGroup:2016lqr}.  Currently, the most compelling proposals to unveil DM particle properties mainly rely on (in-)direct capture or reproduction of DM particles with an array of experiments such as those reported in \cite{XENON100:2013ele} \cite{LUX:2016sci} \cite{PandaX-II:2016vec} \cite{PandaX-II:2016wea} \cite{Gordon:2013vta} \cite{Fermi-LAT:2015att} \cite{Madhavacheril:2013cna} \cite{Bergstrom:2013jra} \cite{Yuan:2017ysv} \cite{Profumo:2017obk} \cite{CMS:2014jvv} \cite{Flauger:2017ged}.  The tiny cross-section, however, may hinder the detection of DM particles.

In this work, to trace the unknown particle nature of DM, we investigate the imprint of DM pair productions encoded in the CMB.  We adopt the NEQDM model \cite{Li:2014era}, which belongs to the broader freeze-in scenario \cite{Baer:2014eja} \cite{Chung:1998ua} \cite{Shi:1998km} \cite{Lin:2000qq} \cite{Feng:2003xh} \cite{Hall:2009bx} \cite{Cheung:2011nn} \cite{Klasen:2013ypa} \cite{Feldstein:2013uha} \cite{Cheung:2014nxi} \cite{Cheung:2014pea}, to study the co-evolution of DM density perturbation and primordial metric perturbation during the DM production phase.  In the NEQDM model, DM particles $\chi$ are produced by pair annihilations of lighter scalar particles $\phi$ ($m_\phi\ll m_\chi$) with the minimal coupling, $\mathcal{L}_{int}=\lambda\phi^2\chi^2$, where we use $\phi$ and $\lambda$ to mimic the standard model (SM) particles and their coupling to DM particles.  The NEQDM model has a small $\lambda$ and can not attain thermal equilibrium. 

We derive the perturbative Einstein and Boltzmann equations, which contain a new driving force term during the DM production phase.  By solving them analytically, we find an amplified perturbative resonance between DM density perturbation and the scalar modes of metric perturbation driven by DM production, which our numerical simulation confirms. 

\begin{figure}
  \includegraphics[width=0.45\textwidth]{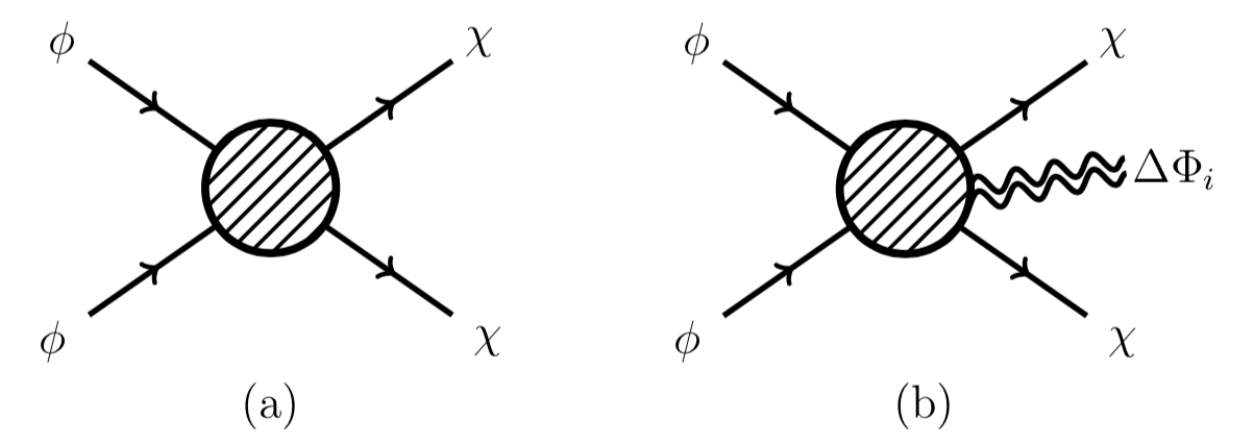}
\caption{An illustration of the pair production of DM particles $\chi$ via a pair annihilation of scalar particles $\phi$ in a flat and rigid spacetime (a. collider physics) and in a curved and nonrigid spacetime (b. cosmology). (b) illustrates a small local metric fluctuation, $\Delta\Phi_i$, generated by a pair production of DM particles in the curved and non-rigid spacetime, which can accumulate to the background metric perturbation during DM production.}
\label{fig:feyman.pdf}       
\end{figure}

In Fig.\ref{fig:feyman.pdf}, we plot the pair productions of DM particles in flat and rigid spacetime (Fig.\ref{fig:feyman.pdf} (a)) and in curved and non-rigid spacetime  (Fig.\ref{fig:feyman.pdf} (b)), respectively, to illustrate the microscopic process of this perturbative resonance.  In contrast to the case shown in Fig.\ref{fig:feyman.pdf} (a) (in collider physics), a small local metric fluctuation, $\Delta\Phi_i$, can be generated when a pair of DM particles are produced in the curved and non-rigid spacetime, which reflects a portion of energy transformed locally from the particles to the metric perturbation in each collision.

In the thermal equilibrium, such extra fluctuation from DM production is canceled precisely by its counterpart from DM annihilation on the interaction-by-interaction basis, $\sum_i \Delta \Phi_i=0$, and no extra fluctuations accumulate. Notably, these extra fluctuations can accumulate during the DM production phase since the DM pair productions dominate over their annihilations. Consequently, these fluctuations accumulate and contribute locally to the scalar modes of background metric perturbation, enhancing them from the primordial value. This local enhancement amplifies the DM density perturbation locally via the back-reaction. Such positive feedback processes thus lead to an amplified perturbative resonance. Note that such perturbative resonance results from the interplay between the background metric perturbation and DM density perturbation. If lacking one of them, this process won't happen. For example, Ref.\cite{Li:2015egy} ignores the metric perturbation for simplicity, so no perturbative resonance occurs in their analysis. 

This perturbative resonance only amplifies the scalar modes of metric perturbation (and DM density perturbation). It does not affect the tensor modes (The extra fluctuations only contribute to the trace components of the metric matrix and do not affect off-diagonal parts.). Therefore, it suppresses the tensor-to-scalar ratio of metric perturbation $r$ from the primordial value. This fact implies that the coming primordial gravitational wave (PGW) searches can examine the perturbative resonance. If future observations confirm suppression of $r$, it could favor a significant perturbative resonance. Adopting a typical cosmic reheating process to specify the cosmic background, we establish a direct relation between $m_\chi$ and $r$, which can constrain $m_\chi$ in future PGW searches. In principle, our proposal based on perturbative resonance is complementary to many other existing strategies primarily based on background DM abundance \cite{ParticleDataGroup:2018ovx}.

\section{The Perturbative Boltzmann and Einstein Equations During DM production} 

In NEQDM model, we adopt the perturbative Friedmann-Lemaitre-Robertson-Walker (FLRW) metric in conformal Newtonian gauge, $ g_{\mu\nu}=\{-1-2\Psi(\vec{x},t), a^2(t)\delta_{ij}$ $\left[1+2\Phi(\vec{x},t)\right]\}$ \cite{Dodelson:2003ft}, to expand the unintegrated Boltzmann equation, $df/dt=C[f]$, to the zeroth order, 
\begin{equation}\label{eq:zoe}
\frac{d n_\chi}{dt}+3Hn_\chi=\widetilde{\langle \sigma v\rangle}\left[\left(n_\phi/n_\phi^{eq}\right)^2\left(n_\chi^{eq}\right)^2-n_\chi^2\right],
\end{equation}
and the first order,
\begin{equation} \label{eq:dtheta}
\frac{d \Theta}{dy}-3\frac{d\Phi}{dy}=\frac{\widetilde{\langle \sigma v \rangle}}{ H y n_\chi}\left[\left(\frac{n_\phi}{n_\phi^{eq}}\right)^2\left(n_\chi^{eq}\right)^2-n_\chi^2\right]\left(\Theta+\Phi\right),
\end{equation}
where $f=\exp[(\mu-E)/T]\left[1-\Theta(\vec{x}, t)\right]$ being the perturbative distribution function of particles, $C$ being the collision term which value determined by $\lambda$; the superscript ${~}^{eq}$ denoting thermal equilibrium; $H$ being the Hubble parameter, $\widetilde{\langle \sigma v \rangle}$  the thermally averaged cross-section, $n_\chi$ and $n_\phi$ the number density of $\chi$ and $\phi$ respectively; $y\equiv m_\chi/T$, $dy/dt=Hy$, $T$ the background temperature in the radiation-dominated epoch; $\Phi(y)$, $\Psi(y)$ and $\Theta(y)$ the long wavelength Fourier modes of $\Phi(\vec{x}, t)$, $\Psi(\vec{x}, t)$ and $\Theta(\vec{x},t)$ respectively. For simplicity, we neglected the terms with $\partial/\partial x^i$ and used $\Phi(y)=-\Psi(y)$ in this work and left the short wavelength modes for further studies. 

\begin{figure}
  \includegraphics[width=0.45\textwidth]{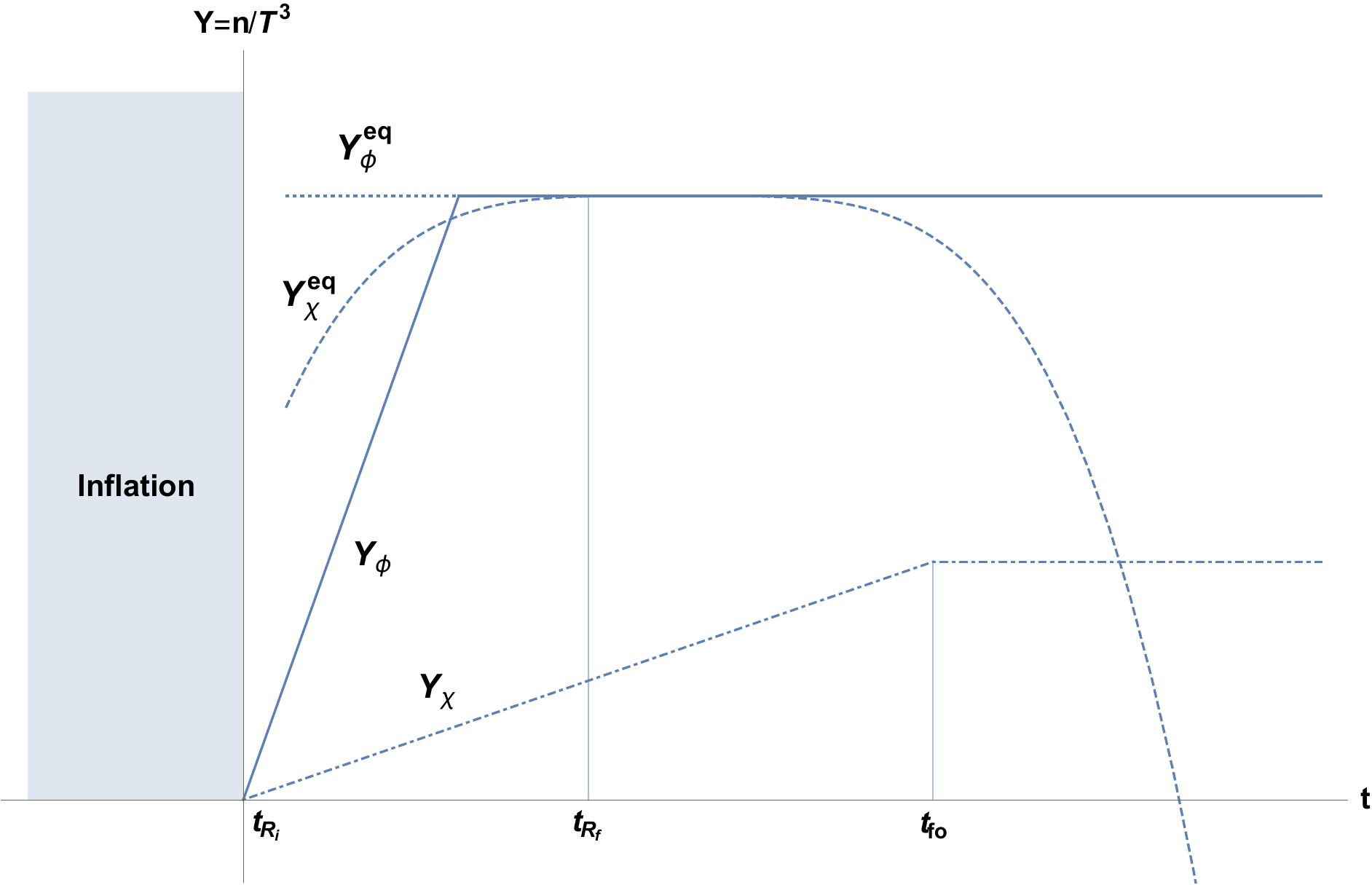}
\caption{A schematic plot of the cosmic evolution of $\chi$ and $\phi$ abundances in the NEQDM model. After inflation, the Universe reheats ($t_{R_i}\le t\le t_{R_f}$) and attains the highest temperature $T_{R_f}=T(t_{R_f})$. $\phi$ abundance (the solid sloped line) attains and immediately tracks its thermal equilibrium (the dotted horizontal line). Due to the tiny cross-section, $\chi$ abundance (the sloped dash-dotted line) increases slowly, can not attain its thermal equilibrium (the dashed curve), and freezes in at $t=t_{fo}$(the horizontal dash-dotted line). $t_{R_f}<t<t_{fo}$ denotes the post-reheating production.}
\label{fig:capab.pdf}       
\end{figure}

The right-hand side (RHS) term in Eq.(\ref{eq:zoe}) (and in Eq.(\ref{eq:dtheta})), which arises from the chemical potential difference of $\phi$ and $\chi$, indicates that both $\chi$ and $\phi$ can be out of thermal equilibrium. Using Eq.(\ref{eq:zoe}), we schematically plot Fig.\ref{fig:capab.pdf} to illustrate the cosmic evolution of the $\chi$ and $\phi$ particles abundances ($Y_\chi\equiv n_\chi/T^3$ and $Y_\phi\equiv n_\phi/T^3$) in the NEQDM model. We assume that at the end of inflation ($t=t_{R_i}$), their initial abundances are zero, $n_\chi(t_{R_i})=0$ and $n_\phi(t_{R_i})=0$. At $t=t_{R_f}$, the Universe is fully thermalized and attains the maximal temperature $T=T_{R_f}$. Compared to $\chi$, $\phi$ has a smaller particle mass and larger total cross-section (with other SM particles), so $\phi$ can attain thermal equilibrium swiftly, which can be even earlier than the cosmic reheating completion ($t=t_{R_f}$) \cite{Bassett:2005xm} \cite{Allahverdi:2010xz} \cite{Amin:2014eta}. On the other hand, the DM particle production process takes a very long time ($t_{f_o}\gg t_{R_f}$) due to the tiny cross-section (a small value of $\lambda$). In the NEQDM model, DM won't attain thermal equilibrium and directly freezes in at  $T(t_{f_o})=m_\chi$. 

Note that in the standard WIMP paradigm, $\phi$ and $\chi$ are assumed in thermal equilibrium initially ($n_\chi=n_\chi^{eq}$ and $n_\phi=n_\phi^{eq}$ at $t=t_{R_i}$), which vanishes the RHS terms in Eq.(\ref{eq:zoe}) and Eq.(\ref{eq:dtheta}) \cite{Dodelson:2003ft}. Notably, the NEQDM model discards this assumption to uncover the DM production process. Therefore, in the NEQDM model, both of the RHS terms in Eq.(\ref{eq:zoe}) and Eq.(\ref{eq:dtheta}) exist. Specifically, the RHS term in Eq.(\ref{eq:zoe}) provides the aforementioned cosmic evolution of $\phi$ and $\chi$ abundance. Furthermore, the RHS term in Eq.(\ref{eq:dtheta}) ({\it i.e.} the perturbative version of the RHS term in Eq.(1)) serves as a driving force and causes a perturbative resonance between the DM density perturbation ($\delta\rho_\chi=-\rho_\chi\Theta$) and the scalar modes of metric perturbation ($\Phi$).

To obtain the complete set of equations of motion for $\Phi$ and $\Theta$, we derive the other equation from the perturbative Einstein equation in the radiation-dominated DM production phase, 
\begin{equation}\label{eq:efo}
\frac{1}{H}\frac{d\Phi}{dt}+\Phi=-\frac{1}{2}\Theta~,
\end{equation} 
which is identical to the conventional expression ({\it c.f.} Eq.(6.6) in Ref.\cite{Dodelson:2003ft}) and reflects a fact that the metric fluctuation $\Phi$ causes the distribution fluctuation $\Theta$.

In the standard WIMP paradigm \cite{Lee:1977ua} \cite{Steigman:1984ac} \cite{Gondolo:1990dk} \cite{Griest:1990kh} \cite{Kolb:1990vq} \cite{Feng:2008ya}, substituting the thermal equilibrium assumption ($n_\chi=n_\chi^{eq}$ and $n_\phi=n_\phi^{eq}$) into Eq.(\ref{eq:dtheta}) and Eq.(\ref{eq:efo}) and using the initial conditions ($\Phi=\Phi_\varphi$ and $d\Phi_\varphi/dt=0$ at $t=t_{R_f}$), one can re-obtain the well-known relation ({\it c.f.}~Eq.(6.12) in Ref.\cite{Dodelson:2003ft}), 
\begin{equation}
    \Theta(y)=-2\Phi(y)=-2\Phi_\varphi~,
\end{equation}
which confirms that the newly derived perturbative equation, Eq.(\ref{eq:dtheta}), is correct. Next, we are solving the background equation Eq.(\ref{eq:zoe}) and the perturbative equations (Eq.(\ref{eq:dtheta}) and Eq.(\ref{eq:efo})) for the NEQDM model to illustrate the perturbative resonance between $\Theta$ and $\Phi$.

\section{Dark Matter Abundance} Using the scaling relation $H=H_my^{-2}$ in the radiation-dominated era and the scaling relation $\widetilde{\langle \sigma v\rangle}=\sigma_0y^{2}/4 $ for bosonic (DM) particles\cite{Li:2014era}, we solve Eq.(\ref{eq:zoe}) with the initial condition ($n_\chi=0$ and $n_\phi=0$ at $t=t_{R_i}$) and obtain
\begin{equation}\label{eq:yin}
Y_\chi(y)=\kappa(y-y_{R_f}+\xi)~,\quad y_{R_f}\le y\le 1~,
\end{equation}
where $\kappa\equiv m_\chi^3\sigma_0(4\pi^4H_m)^{-1}$ and
\begin{equation}\label{eq:xidef}
\xi\equiv \frac{Y_\chi(y_{R_f})}{\kappa}=\frac{4\pi^4H_m}{m_\chi^3\sigma_0T_{R_f}^{3}}\int_{t_{R_i}}^{t_{R_f}}\widetilde{\langle\sigma v\rangle}\left(\frac{n_\chi^{eq}}{n_\phi^{eq}}\right)^2 n_\phi^2 dt.
\end{equation}

In the NEQDM model, the dimensionless parameter $\kappa$ is very small due to the tiny cross-section, and $\kappa\ll (2\pi^2)^{-1}$ is required to ensure the DM not to attain thermal equilibrium (Otherwise, it returns to the standard WIMP paradigm for $\kappa\gg (2\pi^2)^{-1}$.). More specifically, the WIMP paradigm has the relic abundance $Y_f\propto \kappa^{-1}$ as it freezes out, and the NEQDM model has $Y_f\propto \kappa$ as it freezes in.

According to Eq.(\ref{eq:yin}), $Y_\chi(y)$ increases linearly in $y$ . At $y=1$, NEQDM production completes and freezes in the relic abundance, 
\begin{equation}\label{eq:yfi}
    Y_{\chi f}=\kappa(1-y_{R_f}+\xi)\simeq \kappa, \quad y>1,
\end{equation}
with $y_{R_f}\ll \xi \le 1$. Although the newly introduced parameter $\xi$ is negligible in the relic abundance\cite{Li:2014cba}, it encodes the crucial information about the cosmic reheating ($t_{R_i}<t<t_{R_f}$): 1) how the background temperature increases; 2) how $\phi$ produces and attains thermal equilibrium; and 3) how many DM particles produce, respectively, during cosmic reheating and after reheating. Next, we show that this parameter $\xi$ plays a vital role in perturbative resonance.

\section{Perturbative Resonance} 
Substituting Eq.(\ref{eq:efo}) into Eq.(\ref{eq:dtheta}), we obtain
\begin{equation}\label{eq:seoreqfp}
\frac{d^2\Phi}{dy^2}+\frac{7}{2y}\frac{d\Phi}{dy}=\frac{\widetilde{\langle \sigma v \rangle}}{ H y n_\chi}\left[\left(\frac{n_\phi}{n_\phi^{eq}}\right)^2\left(n_\chi^{eq}\right)^2-n_\chi^2\right]\left(\frac{d\Phi}{dy}+\frac{1}{2y}\Phi\right),
\end{equation}
which has a positive driving force term on the RHS. Using the initial condition ($\Phi=\Phi_\varphi$ and $d\Phi(y)/dy=0$ at $y=y_{R_f}$), the background evolutions (Eq.(\ref{eq:yin}) and Eq. (\ref{eq:yfi})) and Eq. (\ref{eq:xidef}) to solve this equation, we obtain 
\begin{equation}\label{eq:phiyev}
\Phi(y)=
\left\{  
  \begin{array} {lr}
 {\displaystyle \Phi_\varphi \mathcal{G}\left(-y\times\xi^{-1}\right),\qquad\quad   y_{R_f}\le y\le 1} 
 \\ 
  {\displaystyle  \Phi_\varphi \mathcal{G}\left(-\xi^{-1}\right)[1+\mathcal{A}(1-y^{-\frac{5}{2}})],  y> 1}   
  \\
\end{array}     
\right.,
\end{equation}
where $\mathcal{G}(x)\equiv {_2F_1}\left(\frac{3-\sqrt{17}}{4}, \frac{3+\sqrt{17}}{4}; \frac{7}{2}; x\right)$ with ${_2F_1}\left(a,b;c,d\right)$ being the Gauss hyper-geometric function, $\mathcal{A}\simeq 0.11$ being obtained by matching the solutions at $y=1$, the decaying mode of the solution $y^{-\frac{5}{2}} {_2F_1}\left(\frac{-7-\sqrt{17}}{4}, \frac{-7+\sqrt{17}}{4};\right.$ $\left. -\frac{3}{2}; -y\times\xi^{-1}\right)$ being neglected, and two approximations $-\xi+y_{R_f}\simeq-\xi$ and $1-\kappa^2\simeq 1$ being taken for simplicity. To cross-check Eq.(\ref{eq:phiyev}), we perform a numerical simulation, reproducing the same result, see Fig.\ref{fig:rxix.pdf}. 

\begin{figure}
  \includegraphics[width=0.45\textwidth]{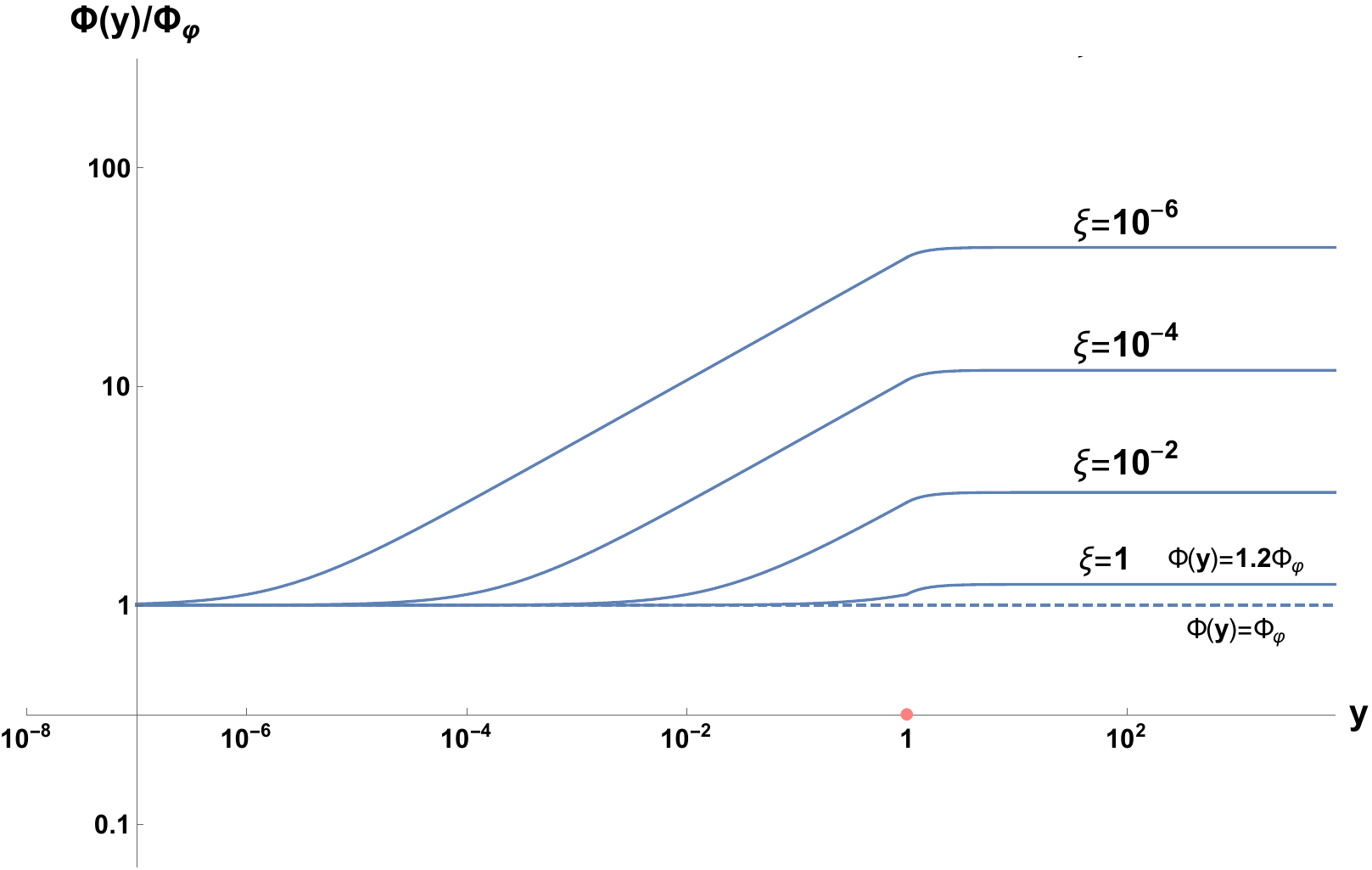}
\caption{The evolution of $\Phi$ for $\xi=(1, 10^{-2}, 10^{-4}, 10^{-6} )$ during the post-reheating epoch $10^{-7}\le y\le 10^{3}$. The dashed line ($\Phi(y)/\Phi_\varphi=1$) corresponds to the standard WIMP paradigm or the NEQDM model with $\xi\gg 1$. }
\label{fig:Phi.pdf}       
\end{figure}

Remarkably, the parameter $\xi$ encodes the cosmic reheating information ($m_\chi$ and $\widetilde{\langle \sigma v \rangle}$) and solely determines the amplitude of perturbative resonance. Using Eq.(\ref{eq:phiyev}), we illustrate the perturbative resonance in $\Phi (y)$ part in Fig.\ref{fig:Phi.pdf}, where we taking $y_{R_f}=10^{-7}$, $10^{-7}\le y\le 10^{3}$ and $\xi=(1, 10^{-2}, 10^{-4}, 10^{-6} )$. (Using Eq.(\ref{eq:efo}), one can obtain $\Theta(y)$ part and DM density perturbation $\delta \rho_\chi=-\rho_\chi \Theta(y)$ accordingly.). We find that a smaller $\xi$ corresponds to a larger perturbative resonance amplitude ($\Phi(y)/\Phi_\varphi$). Using Eq.(\ref{eq:yin}) and Eq.(\ref{eq:yfi}), we can rewrite $\xi$ to interpret it,
\begin{equation}\label{eq:newfoxi}
\xi=Y_\chi(y_{R_f})/Y_{\chi f}~. 
\end{equation} 
This expression implies that $\xi$ reflects the ratio of DM abundance produced during the cosmic reheating and the post-reheating era. Specifically, a smaller $\xi$ corresponds to more DM particles produced after reheating ($t>t_{R_f}$), leading to a larger extra fluctuation accumulation from DM pair productions to drive a stronger perturbative resonance with a larger amplitude amplification. Furthermore, after $y=1$, although DM production completes, the chemical potential pressure between $n_\chi$ and $n_\chi^{eq}$ still exits. Thus it contributes $\mathcal{A}=0.11=11\%$ amplification to the final value of $\Phi(y)$ at $y\rightarrow\infty$.  

Remarkably, our result also applies to the standard WIMP paradigm \cite{Lee:1977ua}, which assumes that DM particles produce instantly and almost no DM particle produces after reheating. Therefore, in the standard WIMP paradigm, $\xi_{WIMP}\gg 1$ and no amplification happens (see the dashed line in Fig.\ref{fig:Phi.pdf}.).  

\section{Scale-invariance and Consistence Correlation.} There are two important relations in the standard model of cosmology\cite{Dodelson:2003ft} \cite{Frenk:2012ph}, the (nearly) scale-invariance of $\Phi_\varphi$ ($d\ln \mathcal{P}_{\Phi_\varphi}/d\ln k\simeq 0$) and the consistency correlation between $\Phi_\varphi$ and DM density perturbation $\delta\rho_\chi=-\rho_\chi\Theta_\varphi$ ($\delta\rho_\chi/\rho_\chi=2 \Phi_\varphi$). The newly unveiled perturbative resonance also preserves these two relations: 1) according to Eq.(\ref{eq:phiyev}), $\Phi(y)/\Phi_\varphi|_{y\gg 1}=\left(-\xi^{-1}\right)[1+\mathcal{A}]$ is independent of the wave-vector $k$, so the scale-invariance preserves ($d\ln \mathcal{P}_{\Phi(y)}/d\ln k\simeq 0$); and 2) using Eq.(\ref{eq:thephi}), we obtain
\begin{equation}\label{eq:thephi}
\left.\frac{\Theta(y)}{\Theta(y_{R_f})}\right|_{y\gg 1}=\left.\left(\frac{y}{\Phi_\varphi}\frac{d\Phi(y)}{dy}+\frac{\Phi(y)}{\Phi_\varphi}\right)\right|_{y\gg 1}=\left.\frac{\Phi(y)}{\Phi_\varphi}\right|_{y\gg 1}~,
\end{equation}
which leads to the consistency correlation,
\begin{equation}
\delta\rho_\chi/\rho_\chi=2 \Phi_\varphi \mathcal{G}\left(-\xi^{-1}\right)[1+\mathcal{A}]=2 \Phi(y)~, \quad y\gg 1~.
\end{equation} 
Note that although such preservation is desired for maintaining the main predictions of the standard model of cosmology, it also prevents one from tracing the perturbative resonance from the discrepancy between $\Phi$ and $\Theta$ (CMB v.s. large-scale surveys on matter distribution).

\section{The Suppression of Tensor-to-Scalar Ratio} The perturbative resonance only amplifies the scalar modes $\Phi$ and does not affects the tensor modes of metric perturbation $h$ (PGW). Therefore, it suppresses the tensor-to-scalar ratio of metric perturbation, 
\begin{equation}\label{eq:ryev}
r\equiv \frac{\mathcal{P}_h}{\mathcal{P}_{\Phi(y)}}=\frac{\mathcal{P}_h}{\mathcal{P}_{\Phi_\varphi}} \times \left(\frac{\Phi_\varphi}{\Phi(y)}\right)^2=9\epsilon\times\left(\frac{\Phi_\varphi}{\Phi(y)}\right)^2~,
\end{equation}
where $\mathcal{P}_{\Phi_\varphi}=\frac{8\pi G}{9k^3}\frac{H^2}{\epsilon}$ and $\mathcal{P}_h=\frac{8\pi G}{k^3}H^2$ being the standard primordial scalar and tensor spectra at the end of cosmic reheating ($y=y_{R_f}$) respectively\cite{Dodelson:2003ft}; $\epsilon\equiv d(H^{-1})/dt$ being the slow-roll parameter of standard inflation, and $r_{standard}\equiv\mathcal{P}_h/\mathcal{P}_{\Phi_\varphi}=9\epsilon$ being the standard (initial) value of $r$ in the standard WIMP paradigm (in the NEQDM model). Using Eq.(\ref{eq:phiyev}), we obtain,
\begin{equation}\label{eq:conxir}
r=7.27\epsilon\times\left[\mathcal{G}\left(-\xi^{-1}\right)\right]^{-2}.
\end{equation}  

\begin{figure}
  \includegraphics[width=0.45\textwidth]{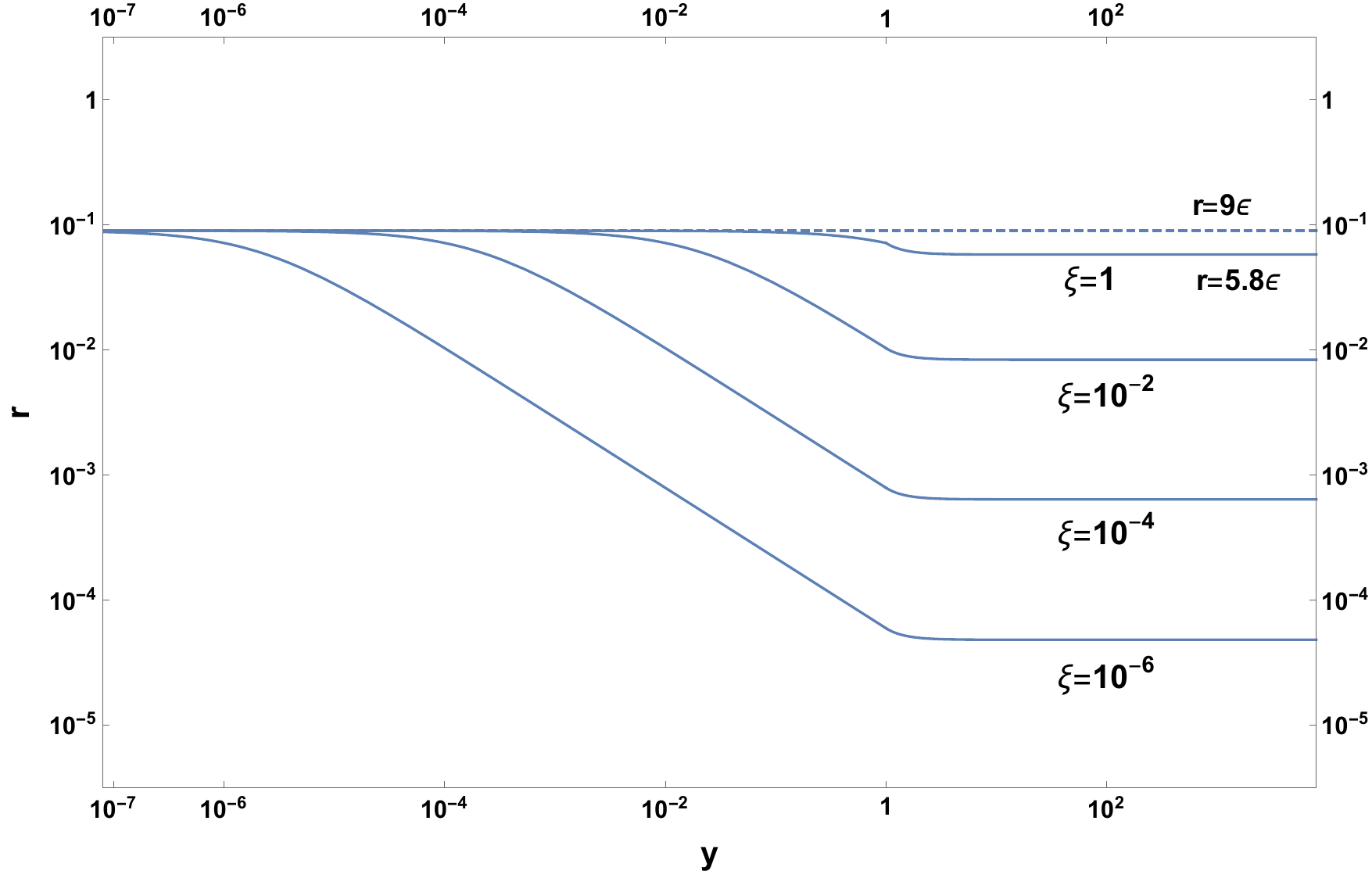}
\caption{The suppression of $r$ for $\xi=(1, 10^{-2}, 10^{-4}, 10^{-6})$ during $10^{-7}\le y\le 1$ with $\epsilon \simeq(1-n_s)/4 \simeq 0.01$ and $n_s\simeq 0.96$. The dashed line ($r=9\epsilon$) corresponds to the standard WIMP paradigm or the NEQDM model with $\xi\gg 1$. }
\label{fig:r.pdf}       
\end{figure}

In FIG.\ref{fig:r.pdf}, we illustrate the suppression of $r$ for $\xi=(1, 10^{-2}, 10^{-4}, 10^{-6})$ by using Eq.(\ref{eq:phiyev}) and Eq.(\ref{eq:ryev}), where the slow-roll parameter and the spectral index taking the standard values $\epsilon \simeq(1-n_s)/4 \simeq 0.01$ and $n_s\simeq 0.96$ respectively\cite{Planck:2015fie} \cite{WMAP:2010qai}, and $10^{-7}\le y\le 1$. Accordingly, a smaller $\xi$ causes a more significant suppression of $r$. And for a very large $\xi$, it returns to the well-known prediction of the standard WIMP paradigm $r=9\epsilon$ (see the dashed line in Fig.\ref{fig:r.pdf}). If the future PGW (the B-modes polarization in CMB) searches probe such a suppression in $r$, it would favor a significant perturbative resonance. 

\section{A direct relation between $m_\chi$ and $r$.} For illustration, we adopt a simple cosmic reheating model \cite{Bassett:2005xm} \cite{Allahverdi:2010xz} \cite{Amin:2014eta}, which background temperature increases linearly during the reheating process ($t_{R_i}\le t\le t_{R_f}$),
\begin{equation}\label{eq:reheo}
T(t)=T_{R_f}\left(\frac{t-t_{R_i}}{t_{R_f}-t_{R_i}}\right)~,
\end{equation} 
to derive the relation of $m_\chi$ and $r$, where $T_{R_f}$ being the highest temperature of the Universe with $T_{R_f}\gg m_\chi$. The duration of reheating can be parameterized as \cite{Kurkela:2011ti} 
\begin{equation}\label{eq:rehet}
t_{R_f}-t_{R_i}\equiv\alpha^{-2}T_{R_f}^{-1}.
\end{equation}
where a larger dimensionless parameter $\alpha$ corresponds to a shorter reheating process. Integrating Eq.(\ref{eq:xidef}) with Eq.(\ref{eq:reheo}) and Eq.(\ref{eq:rehet}), we obtain 
\begin{equation}\label{eq:xicrexp}
\xi=\frac{H_m}{m_\chi T_{R_f}^3}\int^{t_{R_f}}_{t_{m_\chi}}\left[T(t)\right]^4dt=\frac{\pi}{5\alpha^2M_p} m_\chi~,
\end{equation} 
where the reduced Hubble parameter taking $H_m\equiv Hy^2=\pi m_\chi^2/M_p$, and $t_{m_\chi}$ denoting $T=m_\chi$ and reflecting that $\phi$ particles can not produce $\chi$ particles before $t=t_{m_\chi}$. Substituting Eq.(\ref{eq:xicrexp}) into Eq.(\ref{eq:conxir}), we eventually obtain a direct relation between $m_\chi$ and $r$,
\begin{equation}\label{eq:conrm}
r=7.27\epsilon\times\left[\mathcal{G}\left(-5\alpha^2M_p\pi^{-1}m_\chi^{-1}\right)\right]^{-2}~.
\end{equation}
This newly established relation indicates that, in principle, the future PGW (the B-modes in CMB or $r$) searches could directly constrain DM particle mass $m_\chi$. 

\begin{figure}
  \includegraphics[width=0.45\textwidth]{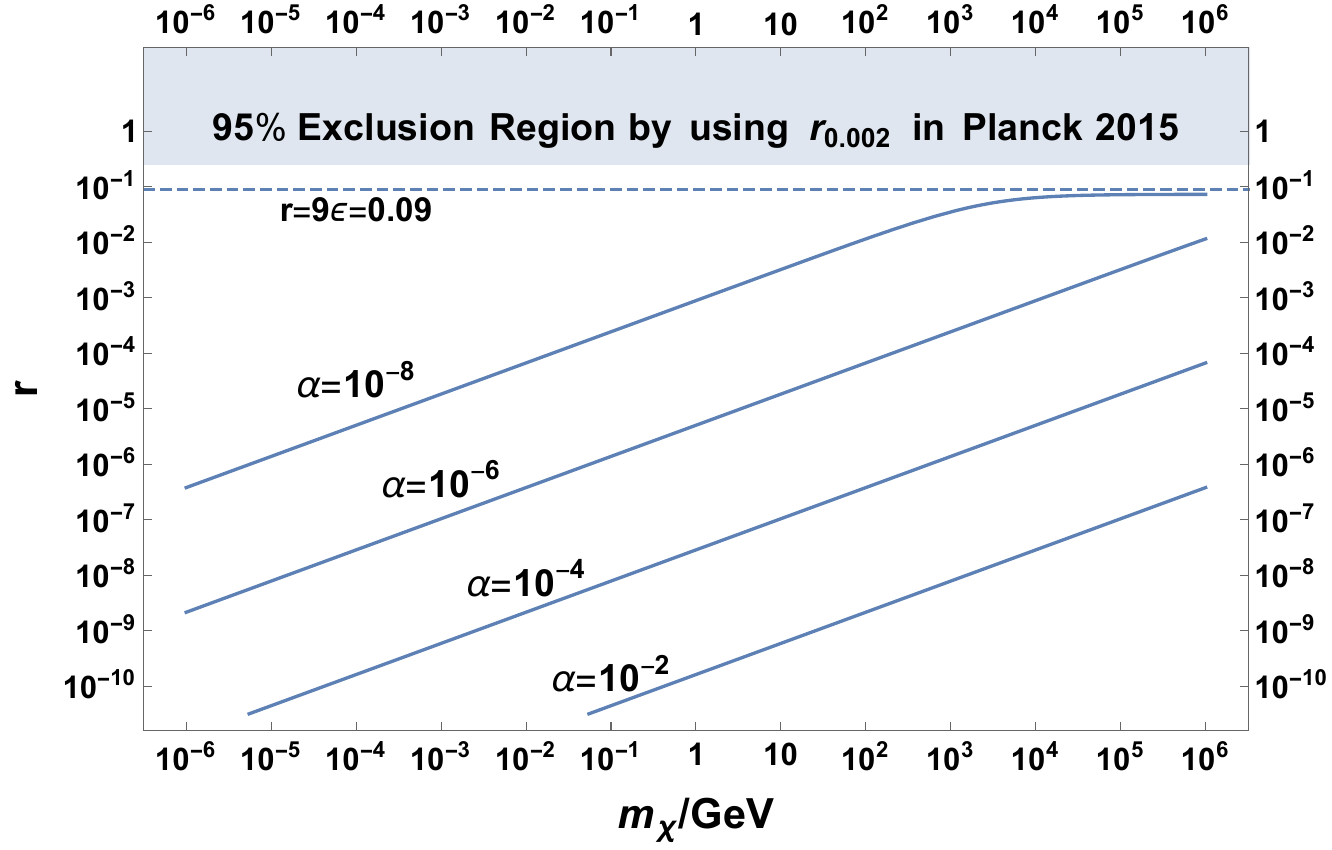}
\caption{The relation between $m_\chi$ and $r$ for  $\alpha=(10^{-2},10^{-4}, 10^{-6}, 10^{-8})$ . }
\label{fig:rmchi.pdf}       
\end{figure}

In FIG.\ref{fig:rmchi.pdf}, we illustrate the relation of $m_\chi$ and $r$ for $\alpha = (10^{-2}, 10^{-4}, 10^{-6}, 10^{-8} )$ by using Eq.(\ref{eq:conrm}) (Note that the parameter space of $m_\chi$ (and $m_\phi$) should be subject to other astrophysical or collider physics constraints, which this work neglects.). The shaded region is the current observational limit on $r$ from Planck satellite\cite{Planck:2015fie}. For a sizeable parameter region, the perturbative resonance suppresses $r$ to be much smaller than the conventional expectation (the dashed line), $r\ll r_{standard}=9\epsilon$, obtained in the standard WIMP paradigm.
We note two features in this plot: 1) A larger $\alpha$ corresponds to a shorter reheating process and more post-reheating DM pair productions; therefore, it causes a stronger suppression of $r$; 2) A smaller $m_\chi$ corresponds to more post-reheating DM pair productions and, again, a stronger suppression of $r$. Accordingly, for a long reheating and a heavy DM candidate ($\alpha=10^{-8}$ and $m_\chi\ge 10^3 \textbf{GeV}$), it returns to the standard prediction $r\rightarrow r_{standard}= 9\epsilon$. Although both conventional and this new prediction of $r$ are beyond current constraints \cite{BICEP2:2014owc} \cite{Planck:2015fie} \cite{BICEP2:2018kqh}, we expect that the coming round PGW searches will distinguish them \cite{CMB-S4:2016ple}. 

\section{Conclusion} 
In this work, adopting the NEQDM model, we unveil a perturbative resonance driven by DM production. When DM production is instant, our proposal reproduces the conventional predictions from the standard WIMP paradigm. However, when DM production lasts longer than the cosmic reheating process (the case in the NEQDM model), such a perturbative resonance can be significant in amplifying the DM density perturbations and the scalar modes of metric perturbations and suppresses the tensor-to-scalar ratio of metric perturbation $r$, which our numerical simulation confirms. Using a simple cosmic reheating model, we obtain a direct relation between $r$ and $m_\chi$, which can be used to constrain $m_\chi$ in future PGW (the B-modes in CMB) searches. In particular, for a sizeable parameter space, the $r$ predicted in our proposal is smaller than the conventional expectation. Therefore, a smaller $r$ measured future could favor our proposal. Our proposal is complementary to many other existing proposals, mainly based on the background DM abundance (for a concise review, see the DM section in Ref.\cite{ParticleDataGroup:2018ovx}).

We highlight several aspects of our proposal for future studies.
\begin{enumerate} 
\item {\it The energy scale of inflation.} The energy scale of inflation is still an open question \cite{Abazajian:2013vfg}. In generic, taking account into the perturbative resonance with a smaller $r$ allows a lower energy scale for inflation.  

\item {\it Cosmic reheating model.} We adopt a simple cosmic reheating model for illustration in this work. However, it is worthwhile to take a realistic cosmic reheating model in the future study\cite{Martin:2014nya}. 

\item {\it The short wavelength modes of DM density perturbation.}  The shorter wavelength modes can re-enter the horizon earlier, get less amplification, and have smaller amplitude. Therefore, it is interesting to relate this issue to the well-known small-scale crisis on a sub-galactic scale\cite{Li:2019yyx}. Hopefully, it can cross-check the amplitude of perturbative resonance, besides the strategy  based on the PGW searches. 

\item {\it Applies to the WIMP paradigm.} The methodology in this study can straightforwardly apply to the WIMP paradigm once taking into account the DM production process of the WIMP paradigm  \cite{Li:2019std}.
\end{enumerate}

In the end, we want to emphasize that the newly unveiled perturbative resonance is not a monopoly of our most straightforward setup (the standard inflation, the NEQDM model with the minimal coupling, and a toy cosmic reheating model) and is worthy of further study in a broader scenario (of various inflation/bounce + DM + reheating models).



\begin{acknowledgements}
We thank Yeuk-Kwan Edna Cheung, Xiaheng Xie, Yajun Wei, and Qing Chen for the beneficial discussion, for carefully reading this manuscript, and suggestions for improving the presentation's clarity. This work has been supported in parts by the National Natural Science Foundation of China (11603018, 11963005, 11775110, 11433004, and 11690030) and Yunnan Provincial Foundation (2016FD006, 2015HA022, and 2015HA030).
\end{acknowledgements}

\section{Supplementary Materials: Numerical Results}

\begin{figure}
  \includegraphics[width=0.45\textwidth]{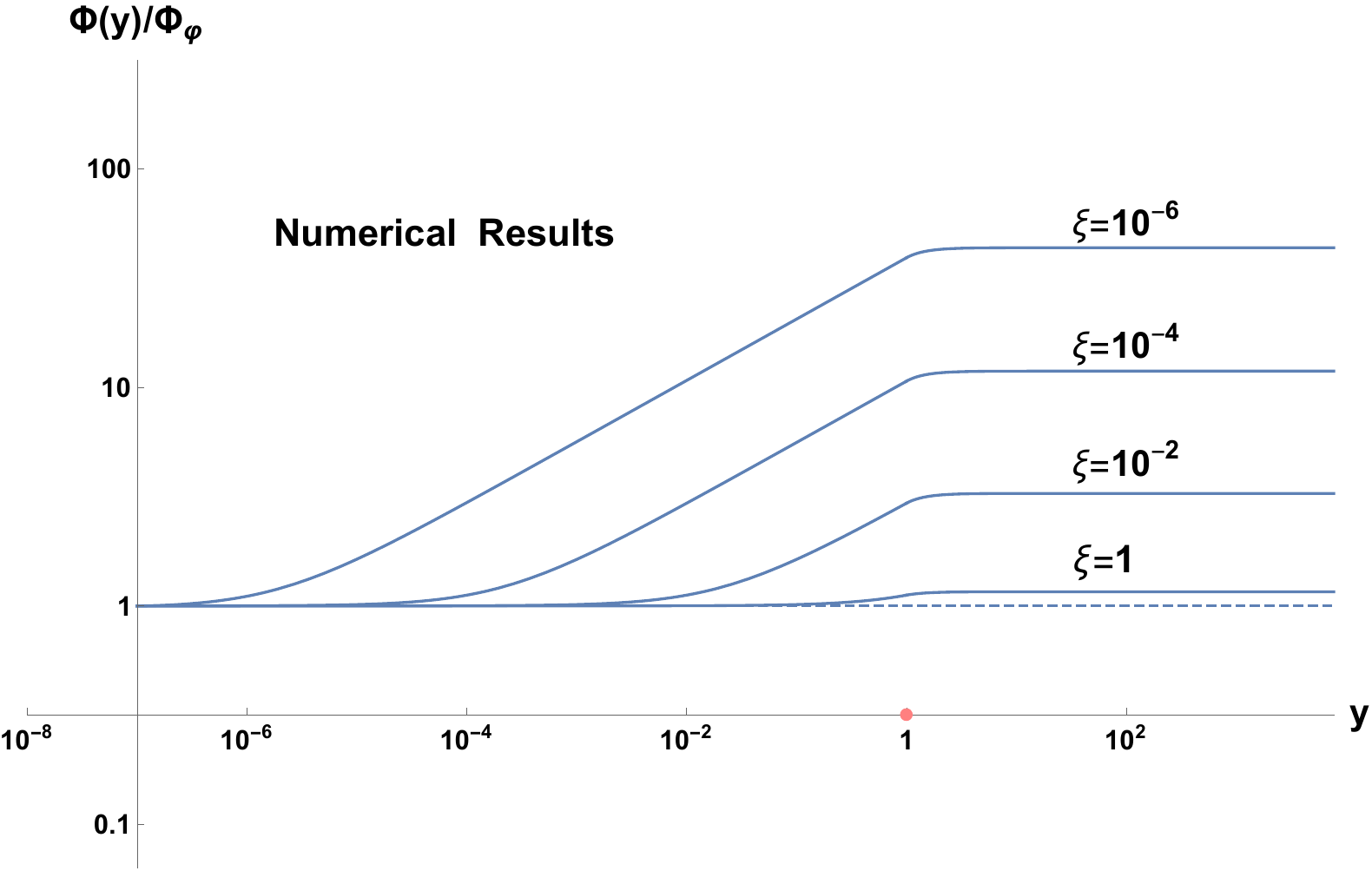}
\caption{The numerical result of the $\Phi$ evolution obtained by solving Eq.(\ref{eq:seoreqfp}) straightforwardly, where the approximation $1-\kappa^2\simeq 1$ being taken.}
\label{fig:NDsolveImpx.pdf}       
\end{figure}

In this section, we solve Eq.(\ref{eq:seoreqfp}) numerically to cross-check the analytical solution (\ref{eq:phiyev}). FIG.\ref{fig:NDsolveImpx.pdf}, the numerical result obtained from Eq.(\ref{eq:seoreqfp}) is the same as FIG.\ref{fig:Phi.pdf}, the analytical solution obtained from Eq.(\ref{eq:phiyev}), which confirms our analytical solution.

\begin{figure}
  \includegraphics[width=0.45\textwidth]{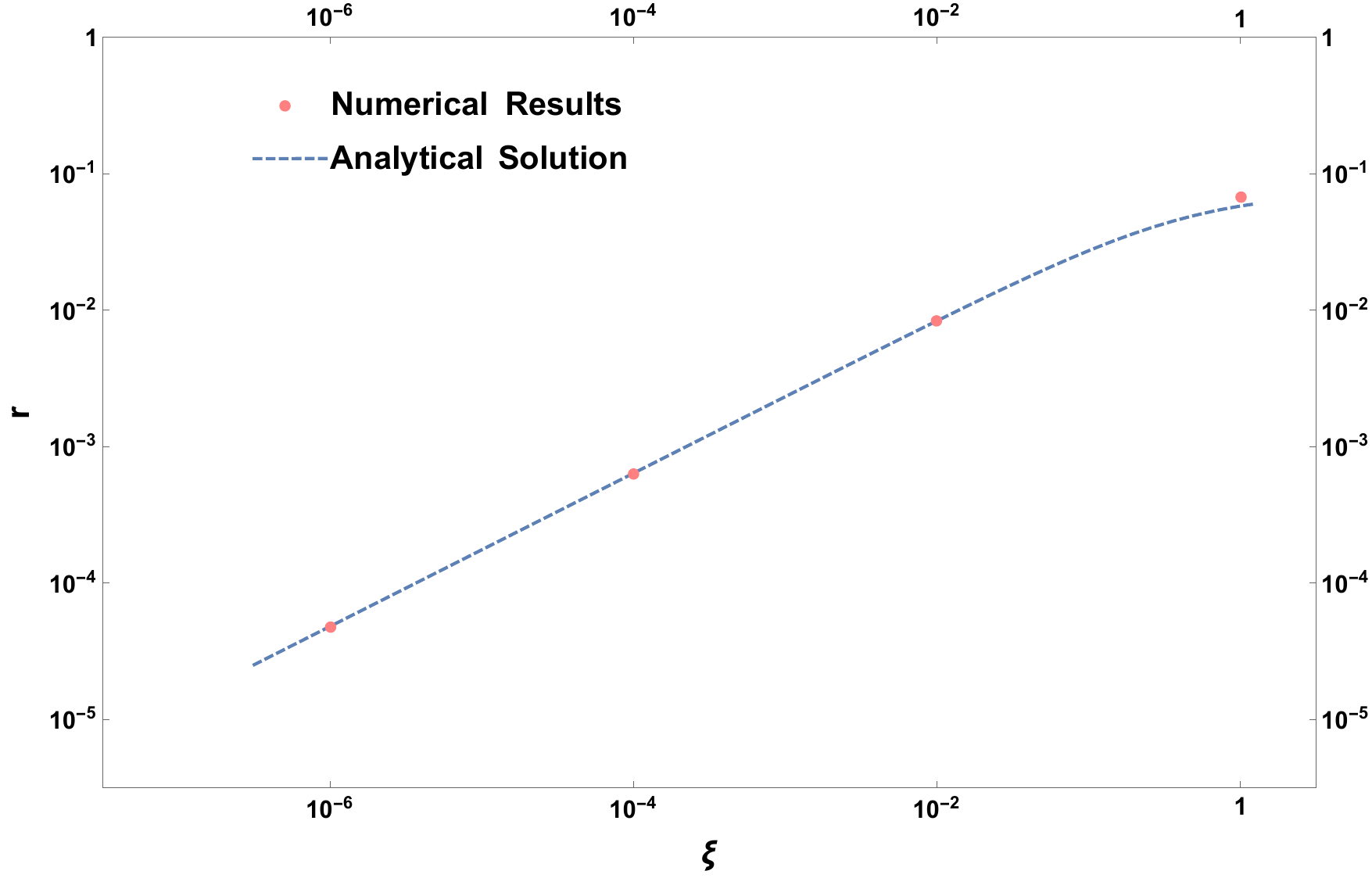}
\caption{The comparison of the numerical and analytical results in $(\xi, r)$ plane. The points of the numerical results are for $\xi={(1, 10^{-2}, 10^{-4}, 10^{-6})}$ respectively. }
\label{fig:rxix.pdf}       
\end{figure}

To make the comparison more apparent, we plot the results from Eq.(\ref{eq:seoreqfp}) and Eq.(\ref{eq:phiyev}) respectively in $(\xi, r)$ plane with $\epsilon=0.01$, as shown in FIG.\ref{fig:rxix.pdf}. They match perfectly. Thus we can safely conclude that the analytical solution, Eq.(\ref{eq:phiyev}), is, mathematically, correct. 

\end{document}